\input epsf

\newfam\msbfam
\font\twlmsb=msbm10 at 12pt
\font\eightmsb=msbm10 at 8pt
\font\sixmsb=msbm10 at 6pt
\textfont\msbfam=\twlmsb
\scriptfont\msbfam=\eightmsb
\scriptscriptfont\msbfam=\sixmsb
\def\cj{\fam\msbfam}

\def\R{{\cj R}}

\centerline{\bf WEYL GEOMETRY, ANTI-DE SITTER SPACE, AND $\Phi^4$-THEORY} 

\

\centerline{H. A. Camargo}
\centerline{\it Facultad de Ciencias, Universidad Nacional Aut\'onoma de M\'exico}
\centerline{\it Circuito Exterior, Ciudad Universitaria, 04510, M\'exico D. F., M\'exico} 

\

\centerline{E. J. Gonzalez de Urreta$^*$}
\centerline{\it Instituto de Ciencias, Universidad Nacional de General Sarmiento}
\centerline{\it Juan Mar\'\i a Guti\'errez 1150 (B1613GSX), Los Polvorines, Pcia. de Buenos Aires, Argentina}

\ 

\centerline{and}

\

\centerline{M. Socolovsky$^{**}$}
\centerline{\it  Instituto de Ciencias Nucleares, Universidad Nacional Aut\'onoma de M\'exico}
\centerline{\it Circuito Exterior, Ciudad Universitaria, 04510, M\'exico D. F., M\'exico}

\

{\bf Abstract.} {\it We study the Gaussian approximation to the quantum fluctuations of the metric of the four dimensional anti-De Sitter spacetime. The associated massless scalar field has a quartic self interaction, for which we construct the generating functional of the $n$-point correlation functions. The concomitant infrared divergence is cured by a mass renormalization provided by the cosmological constant, which is also responsible for the renormalization of the coupling constant of the field theory.}

\

{\bf 1. Introduction}

\

The equivalence of a Weylian geometry with vanishing 1-form or, equivalently, with zero gauge curvature, with a Riemannian geometry, allows us to interpret a conformally flat metric as a massless scalar field in Minkowski space. In particular, we consider the four dimensional anti-De Sitter spacetime, and study the Gaussian approximation to the quantum fluctuations of the metric. This is done through the computation of the propagator between two conformal metrics along a finite interval in one of the two patches of the AdS space. A change of the field variable, however, shows that the system is equivalent to a massless $\Phi^4$-theory in one spacetime dimension. We then construct the generating functional of the $n$-point correlation functions, which  give the full quantum fluctuations of the AdS metric around its classical value. The infrared divergence of the scalar theory, due to its massless character, is cured by a finite mass renormalization provided by the cosmological constant, which plays the role of an IR cutoff; this is already exhibited by the two-point function calculated to first order in the coupling constant, which is also finitely renormalized by the four-point function with the same cutoff, at zero momenta, and to second order in the perturbation expansion.  

\

{\bf 2. Weyl and Riemann geometries}

\

As is well known$^1$ any Riemannian (or pseudo-Riemannian, in particular Lorentzian) manifold $(M_{\cal R}^n,g_{\mu\nu})$ can can be considered as a Weylian manifold with zero Weyl field $\phi$ locally given by the 1-form $\phi=\phi_\mu dx^\mu$, $(M_{\cal W}^n, g_{\mu\nu},0)$, with $M_{\cal R}^n=M_{\cal W}^n\equiv M^n$ as differentiable manifolds. Of particular interest is the case when $M^n$ is {\it Lorentzian} and {\it conformally flat} i.e. $$g_{\mu\nu}=e^\varphi\eta_{\mu\nu} \eqno{(1)}$$ where $\varphi$ is a real valued scalar function on $M^n$ and $\eta_{\mu\nu}$ is the Lorentz metric; so $$(M_{\cal W}^n,g_{\mu\nu},0)=(M^n,e^\varphi\eta_{\mu\nu},0).\eqno{(2)}$$ However, any Weyl manifold belongs to the equivalent class of manifolds $[(M^n,g^\prime_{\mu\nu},\phi^\prime)]$ defined by $$(M^n,g^{\prime\prime}_{\mu\nu},\phi^{\prime\prime}) \sim (M^n,g^\prime_{\mu\nu},\phi^\prime) \eqno{(3)}$$ if and only if $$g^{\prime\prime}_{\mu\nu}=\lambda^{-1}g^\prime_{\mu\nu}\eqno{(4a)}$$ and $$\phi^{\prime\prime}=\phi^\prime-\lambda^{-1}d\lambda, \eqno{(4b)}$$ where $\lambda$ is a smooth positive real valued function defined on $M^n$. (4$a$) is a conformal transformation of the metric, while (4$b$) is a gauge transformation of the Weyl field. Both, (4$a$) and (4$b$), constitute what is called a {\it Weyl transformation}, denoted by {\it w}. A choice of a representative of the class is called a {\it choice of gauge} or {\it gauge fixing}. Locally, the Weyl connection $\Gamma^\alpha_{\nu\mu}$ is given by the sum of the Levi-Civita connection $$\pmatrix{\alpha\cr\nu\mu\cr}={{1}\over{2}}g^{\alpha\beta}(\partial_\nu g_{\mu\beta}+\partial_\mu g_{\nu\beta}-\partial_\beta g_{\nu\mu})\eqno{(5)}$$ and the Weyl tensor $$W^\alpha_{\nu\mu}={{1}\over{2}}(\delta^\alpha_\nu\phi_\mu+\delta^\alpha_\mu\phi_\nu-g_{\nu\mu}g^{\alpha\rho}\phi_\rho).\eqno{(6)}$$ It is easily verified that $$\Gamma^\alpha_{\nu\mu}=\pmatrix{\alpha\cr\nu\mu\cr}+W^\alpha_{\nu\mu}\eqno{(7)}$$ is invariant under {\it w}, and so are the curvature tensor $$R^\rho_{\sigma\mu\nu}=\partial_\mu\Gamma^\rho_{\nu\sigma}-\partial_\nu\Gamma^\rho_{\mu\sigma}+\Gamma^\rho_{\mu\beta}\Gamma^\beta_{\nu\sigma}-\Gamma^\rho_{\nu\beta}\Gamma^\beta_{\mu\sigma}\eqno{(8)}$$ and the Ricci tensor $$R_{\nu\sigma}=R^\mu_{\nu\mu\sigma}.\eqno{(9)}$$ However, since $$g^{\mu\nu}\buildrel{w}\over\longrightarrow g^{\prime\mu\nu}=\lambda g^{\mu\nu},\eqno{(10)}$$ the curvature scalar $$R=g^{\mu\nu}R_{\mu\nu}\eqno{(11)}$$ is not an invariant but transforms to $\lambda R$, i.e. $$R\buildrel{w}\over\longrightarrow R^\prime=\lambda R,\eqno{(12)}$$ leaving the Einstein tensor $$G_{\mu\nu}=R_{\mu\nu}+{{1}\over{2}}g_{\mu\nu}R\eqno{(13)}$$ invariant. 

\

Applying (3), (4$a$) and (4$b$) to (2) with $\phi^\prime=0$ and $\lambda=e^\varphi$ one obtains $$(M^n,e^\varphi\eta_{\mu\nu},0)\sim(M^n,\eta_{\mu\nu},-d\varphi).\eqno{(14)}$$ That is, a conformally flat Lorentzian manifold is equivalent to a flat Weylian manifold $(M^n,\eta_{\mu\nu},\phi^{\prime\prime})$ with Weyl field $$\phi^{\prime\prime}_\mu=-\partial_\mu\varphi\equiv-\varphi_{,\mu}.\eqno{(15)}$$ Since $\phi^{\prime\prime}=-d\varphi$, its curvature vanishes: $$F=d\phi^{\prime\prime}=0.\eqno{(16)}$$

\

So, {\it classically}, the whole gravitational information in $(M^n,e^\varphi\eta_{\mu\nu})$ is encoded in the scalar $\varphi$ living in Minkowski space $Mink^n$.

\

For any constant scalar curvature Riemannian or pseudo-Riemannian manifold, with curvature tensor $$R_{\mu\nu\rho\sigma}={{R}\over{n(n-1)}}(g_{\mu\rho}g_{\nu\sigma}-g_{\mu\sigma}g_{\nu\rho}),\eqno{(17)}$$ it is easy to verify that the Weyl curvature tensor$^2$ $C_{\mu\nu\rho\sigma}$ defined by $$R_{\beta\sigma\mu\nu}=C_{\beta\sigma\mu\nu}-{{R}\over{(n-1)(n-2)}}(g_{\mu\beta}g_{\nu\sigma}-g_{\nu\beta}g_{\mu\sigma})+{{1}\over{(n-2)}}(g_{\mu\beta}R_{\nu\sigma}-g_{\nu\beta}R_{\mu\sigma}-g_{\mu\sigma}R_{\nu\beta}+g_{\nu\sigma}R_{\mu\beta})\eqno{(18)}$$ vanishes everywhere, and, since a necessary and sufficient condition for a metric to be conformally flat is that its Weyl curvature tensor vanishes everywhere$^3$, then all maximal symmetric spaces are conformally flat. 

\

{\bf 3. Anti-De Sitter space}

\

As a concrete example, we consider the anti-de Sitter space in four space-time dimensions ($AdS_4$) whose metric is given by $$ds^2=({{a}\over{z}})^2(dt^2-dy_1^2-dy_2^2-dz^2)=({{a}\over{z}})^2\eta_{\mu\nu}dx^\mu dx^\nu\eqno{(19)}$$ with $x^0=t$, $x^i=y^i$ $\in (-\infty,+\infty)$, $i=1,2$, $x^3=z\in (-\infty,0)\cup(0,+\infty)$ ($(-\infty,0)$ and $(0,+\infty)$ defining the left and right patches of $AdS_4$ respectively), $[t]=[y^i]=[z]=[a]=[length]$, and $a$ related to the attractive cosmological constant $\Lambda$ through $$\Lambda=-{{3}\over{a^2}}.\eqno{(20)}$$ It is clear that from (1) and (19), $e^\varphi=({{a}\over{z}})^2={{a^2}\over{\vert z\vert^2}}$ and therefore $$\varphi\equiv\varphi_0(z)=2 \ ln({{a}\over{\vert z\vert}}),\eqno{(21)}$$ which is plotted in Figure 1:

\

\centerline{\epsfxsize=70ex\epsfbox{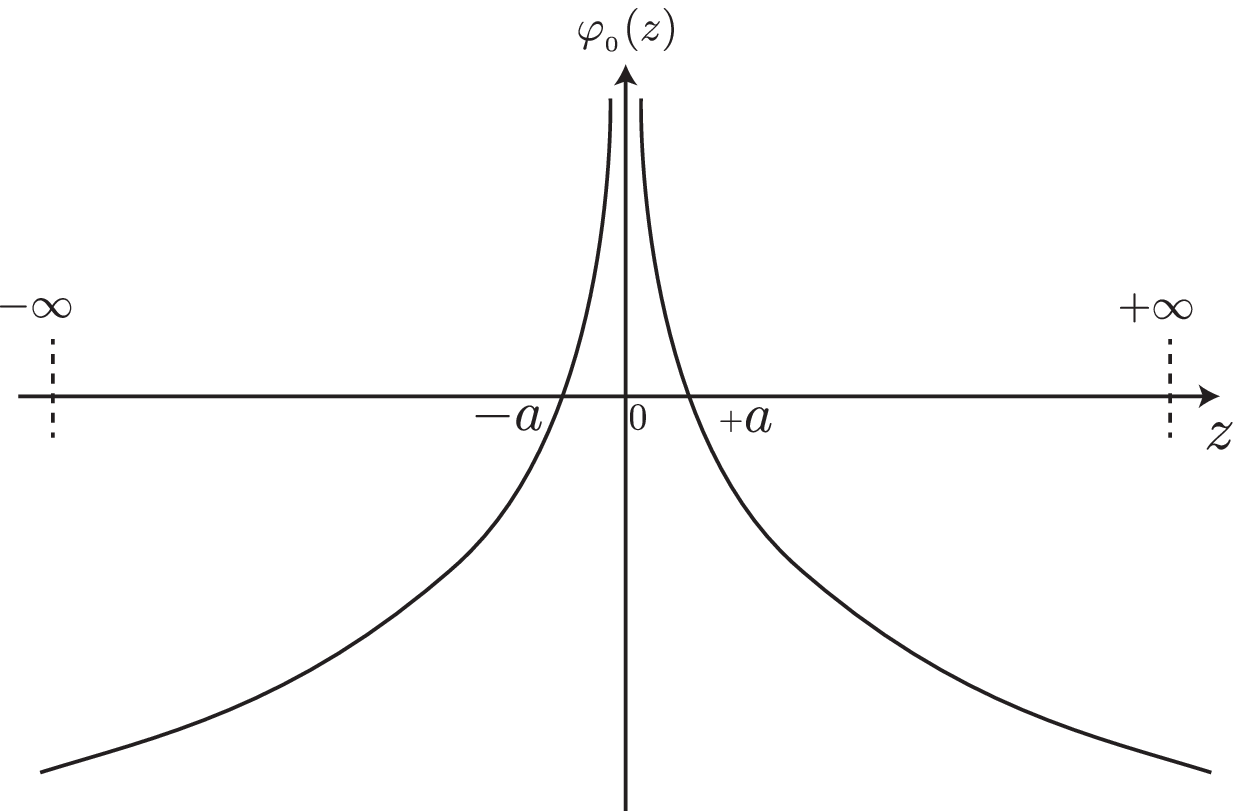}}

\

\centerline{Figure 1}

\

$z=0$ corresponds to the boundary with radial coordinate $r=+\infty$ and $z=\pm\infty$ are the coordinate values of the ``right" and ``left" horizons. 

\

$\varphi_0(z)$ obeys the non-linear partial differential equation $$\eta^{\mu\nu}\partial_\mu\partial_\nu\varphi+{{1}\over{2}}\eta^{\mu\nu}(\partial_\mu\varphi)(\partial_\nu\varphi)+{{4}\over{a^2}}e^\varphi=0, \eqno{(22)}$$ with Lagrangian density $${\cal L}={{1}\over{a^2}}(e^\varphi\eta^{\mu\nu}(\partial_\mu\varphi)(\partial_\nu\varphi)-{{4}\over{a^2}}e^{2\varphi})\eqno{(23)}$$ and action $$S=\int d^4x{\cal L}(z).\eqno{(24)}$$ Equation (22) is obtained from the vacuum Einstein equation for a conformally flat metric $e^\varphi\eta_{\mu\nu}$ in the presence of a cosmological constant $\Lambda$: $$\partial_\mu\partial_\nu\varphi-\eta_{\mu\nu}\eta^{\alpha\beta}\partial_\alpha\partial_\beta\varphi-{{1}\over{2}}(\partial_\mu\varphi)(\partial_\nu\varphi)-{{1}\over{4}}\eta_{\mu\nu}(\partial_\alpha\varphi)(\partial^\alpha\varphi)+\Lambda\eta_{\mu\nu}e^\varphi=0,\eqno{(25)}$$ through its contraction with the inverse Minkowski metric $\eta^{\mu\nu}$.

\

From the statement immediately after eq.(16), the quantum fluctuations of the metric (19) are given by the quantum theory of the field $\varphi$ with action (24) in $Mink^4$. An important element which must then be studied in this context, is the propagator from the conformal metric $g_{\mu\nu}^{(1)}(z)=e^{\varphi_1(z)}\eta_{\mu\nu}$  to the conformal metric $g_{\mu\nu}^{(2)}(z)=e^{\varphi_2(z)}\eta_{\mu\nu}$, with $z_1\leq z\leq z_2$, $0\notin [z_1,z_2]$, $\varphi_i(z)=\varphi_0(z)+\psi_i(z)$, $\psi_i(z_k)=0$, $i=1,2$, $k=1,2$, which is represented by the Feynman path integral over the field $\varphi$, i.e. $$K(g_{\mu\nu}^{(2)},g_{\mu\nu}^{(1)})\equiv K(2,1)=\int_{g_{\mu\nu}^{(1)}}^{g_{\mu\nu}^{(2)}}{\cal D}\varphi e^{iS[\varphi]}.\eqno{(26)}$$ From the decomposition $$\varphi(z)=\varphi_0(z)+\psi(z),\eqno{(27)}$$ (26) becomes $$K(2,1)=\int_{\psi_1}^{\psi_2}{\cal D}\psi e^{i\int_{z_1}^{z_2}d^4x{\cal L}(\varphi_0+\psi)}=\int_{\psi_1}^{\psi_2}{\cal D}\psi e^{-4i\int_{z_1}^{z_2}d^4x{{e^\psi}\over{z^4}}(1-z\psi^\prime+{{z^2}\over{4}}{\psi^\prime}^2+e^\psi)}$$ $$=\int_{\psi_1}^{\psi_2}{\cal D}\psi e^{-4i{\cal V}_\infty\int_{z_1}^{z_2}dz{{e^\psi}\over{z^4}}(1-z\psi^\prime+{{z^2}\over{4}}{\psi^\prime}^2+e^\psi)}, \eqno{(28)}$$ where we assumed that the quantum fluctuation $\psi$ only depends on $z$, with $\psi^\prime={{d\psi}\over{dz}}$, and ${\cal V}_\infty =\int_{-\infty }^{+\infty}dt dx^1dx^2$ is an infinite volume transverse to $z$ which will be considered below.

\

{\bf 4. Semiclassical quantization}

\

The path integral (28) can not be done exactly. So, we consider here its Gaussian or semiclassical (quadratic) approximation, which we call $K_{2}(2,1)$. From $$e^\psi(1-z\psi^\prime+{{z^2}\over{4}}{\psi^\prime}^2)+e^{2\psi}=2+3\psi-z\psi^\prime+{{5}\over{2}}\psi^2-z\psi\psi^\prime+{{z^2}\over{4}}{\psi^\prime}^2+cubic \ terms,\eqno{(29)}$$ we have $$K_2(2,1)=\int_{\psi_1}^{\psi_2}{\cal D}\psi e^{-4i{\cal V}_\infty \int_{z_1}^{z_2}{{dz}\over{z^4}}(2+3\psi-z\psi^\prime+{{5}\over{2}}\psi^2-z\psi\psi^\prime+{{z^2}\over{4}}{\psi^\prime}^2)}=\int_{\psi_1}^{\psi_2}{\cal D}\psi e^{iS}\eqno{(30)}$$ where $$S=-4({{{\cal V}_\infty}\over{a^3}})\int_{z_1}^{z_2}dz{\cal L}_2(z)\eqno{(31)}$$ with $${\cal L}_2(z)={{a^3}\over{z^4}}(2+3\psi-z\psi^\prime+{{5}\over{2}}\psi^2-z\psi\psi^\prime+{{z^2}\over{4}}{\psi^\prime}^2).\eqno{(32)}$$ The Lagrange equation for ${\cal L}_2$, $${{d}\over{dz}}({{\partial{\cal L}_2}\over{\partial\psi^\prime}})-{{\partial{\cal L}_2}\over{\partial\psi}}=0\eqno{(33)}$$ is $$\psi^{\prime\prime}-{{2}\over{z}}\psi^\prime-{{4}\over{z^2}}\psi=0\eqno{(34)}$$ ($\psi^{\prime\prime}={{d^2\psi}\over{dz^2}}$) which, in terms of the dimensionless coordinate $$\hat{z}={{z}\over{a}}\eqno{(35)}$$ becomes $$\ddot{\psi}-{{2}\over{\hat{z}}}\dot{\psi}-{{4}\over{\hat{z}^2}}\psi=0\eqno{(36)}$$ with $\dot{\psi}={{d\psi}\over{d\hat{z}}}$ and $\ddot{\psi}={{d^2\psi}\over{d\hat{z}^2}}$. Its general solution is $$\bar{\psi}(\hat{z})=C_1\hat{z}^4+{{C_2}\over{\hat{z}}}, \ C_1,C_2\in \R.\eqno{(37)}$$ In terms of $z$, $$\bar{\psi}(z)={{C_1}\over{a^4}}z^4+{{C_2a}\over{z}},\eqno{(38)}$$ leading to $$\int_{z_1}^{z_2}{{dz}\over{z^4}}(2+3\bar{\psi}-z\bar{\psi}^\prime+{{5}\over{2}}\bar{\psi}^2-z\bar{\psi}\bar{\psi}^\prime+{{z^2}\over{4}}(\bar{\psi}^\prime)^2)=[-{{2}\over{3z^3}}-{{C_1}\over{a^4}}z-{{C_2a}\over{z^4}}+{{C_1^2}\over{2a^8}}z^5-{{3}\over{4}}{{C_2^2a^2}\over{z^5}}]_{z_1}^{z_2}.\eqno{(39)}$$ Since $\bar{\psi}$ has to obey the same boundary conditions as $\psi$, i.e. $\bar{\psi}(z_1)=\bar{\psi}(z_2)=0$, the unique solution (38) satisfying these conditions with $z_1\neq z_2$ is $\bar{\psi}(z)=0$ i.e. $C_1=C_2=0$. So $$\int_{z_1}^{z_2}{{dz}\over{z^4}}(2+3\bar{\psi}-z\bar{\psi}^\prime+{{5}\over{2}}\bar{\psi}^2-z\bar{\psi}\bar{\psi}^\prime+{{z^2}\over{4}}(\bar{\psi}^\prime)^2)={{2}\over{3}}({{1}\over{z_1^3}}-{{1}\over{z_2^3}}), \eqno{(40)}$$ which gives the action $$\bar{S}(\bar{\psi}=0)=-{{8}\over{3}}({{{\cal V}_\infty}\over{a^3}})a^3({{1}\over{z_1^3}}-{{1}\over{z_2^3}}).\eqno{(41)}$$ ${\cal V}_\infty/a^3$ is a dimensionless infinite transverse volume. Since the problem is essentially 1-dimensional -only the $z$ coordinate is relevant- we define the action per unit dimensionless transverse volume $$\sigma={{\bar{S}}\over{{\cal V}_\infty/a^3}}=-{{8}\over{3}}a^3({{1}\over{z_1^3}}-{{1}\over{z_2^3}}).\eqno{(42)}$$ The van-Vleck determinant$^{4,5}$ associated with $\sigma$ is $$\Delta=det\pmatrix{{{\partial^2\sigma}\over{\partial\hat{z_1}^2}} & {{\partial^2\sigma}\over{\partial\hat{z_1}\partial\hat{z_2}}}\cr {{\partial^2\sigma}\over{\partial\hat{z_2}\partial\hat{z_1}}} & {{\partial^2\sigma}\over{\partial\hat{z_2}^2}}}={{(32)^2a^{10}}\over{(z_1z_2)^5}},\eqno{(43)}$$ which, together with the exponentiation of $\sigma$, gives the desired propagator ($k_2$) for the 1-dimensional problem i.e. that ignoring the irrelevant transverse coordinates $t$, $x^1$ and $x^2$: $$k_2(2,1)=\sqrt{{{i\Delta}\over{2\pi}}}e^{i\sigma}=\sqrt{{{i}\over{2\pi}}}{{32a^5}\over{(z_1z_2)^{5/2}}}e^{-{{8}\over{3}}ia^3({{1}\over{z_1^3}}-{{1}\over{z_2^3}})}.\eqno{(44)}$$

\

{\bf 5. $\Phi^4$-theory}

\

A simple change of field variable, $\varphi\to\Omega$, shows that the Lagrangian (23) corresponds to a $\Phi^4$-theory for the real massless scalar field $\Omega$ in one spacetime dimension -that corresponding to the $z$ coordinate-. In fact, with $$e^\varphi=\Omega^2 \eqno{(45)}$$ i.e. the field $\Omega$ gives the standard form for the conformal factor in (1), the Lagrangian density (23) becomes $${\cal L}={{4}\over{a^2}}(\eta^{\mu\nu}(\partial_\mu\Omega)(\partial_\nu\Omega)-{{1}\over{a^2}}\Omega^4).\eqno{(46)}$$ As for $\varphi$, $[\Omega]=[L]^0$ and $[{\cal L}]=[L]^{-4}$. Assuming again that $\Omega=\Omega(z)$, ${\cal L}$ becomes $${\cal L}=-{{4}\over{a^2}}(({{d\Omega}\over{dz}})^2+{{1}\over{a^2}}\Omega^4)\eqno{(47)}$$ with action $$S=\int d^4x{\cal L}=-4a({{{\cal V}_\infty}\over{a^3}})\int dz(({{d\Omega}\over{dz}})^2+{{1}\over{a^2}}\Omega^4). \eqno{(48)}$$ In terms of the dimensionless coordinate $$\zeta={{z}\over{a}}, \eqno{(49)}$$ $\Omega=\Omega(\zeta)$ and the classical action becomes $$S=-4({{{\cal V}_\infty}\over{a^3}})\int d\zeta(({{d\Omega}\over{d\zeta}})^2+\Omega^4). \eqno{(50)}$$ Again, defining an action per unit dimensionless transverse volume, one obtains $$\sigma=\int d\zeta \ l=-4\int d\zeta(({{d\Omega}\over{d\zeta}})^2+\Omega^4) \eqno{(51)}$$ with Lagrangian density $$l=-4(({{d\Omega}\over{d\zeta}})^2+\Omega^4) \eqno{(52)}$$ and classical equation of motion $$\Delta_\zeta\Omega+2\Omega^3=0, \ \Delta_\zeta=-{{d^2}\over{d\zeta^2}}\eqno{(53)}$$ with solution $$\Omega_0(\zeta)=\zeta^{-1}\to 0 \ \ as \ \ \zeta\to\pm\infty.\eqno{(54)}$$ (cf.(21).)

\

The Feynman ``propagator" is the Green function of the free one dimensional Klein-Gordon equation with a delta function source: $$(\Delta_\zeta-i\epsilon)\Delta_F(\zeta)=-\delta(\zeta),\eqno{(55)}$$namely$^6$ $$\Delta_F(\zeta)=\int_{-\infty}^{+\infty}{{d\xi}\over{2\pi}}{{e^{-i\xi\zeta}}\over{\xi^2+i\epsilon}}.\eqno{(56)}$$ As it stands, $\Delta_F(\zeta)$ is infrared (IR) divergent (as $\xi\to 0$); this is so because the bare mass of the $\Omega$ field is zero. A natural IR cutoff is $$\tilde{\mu}^2=3a^{-2}\eqno{(57)}$$ i.e. minus the cosmological constant $\Lambda$. In dimensionless units, $$\mu^2=a^2\tilde{\mu}^2=3\eqno{(58)}$$ and so the IR finite propagator is $$\Delta_F(\zeta)^\prime=\int_{-\infty}^{+\infty}{{d\xi}\over{2\pi}}{{e^{-i\xi\zeta}}\over{\xi^2-\mu^2+i\epsilon}}=\int_{-\infty}^{+\infty}{{d\xi}\over{2\pi}}{{e^{-i\xi\zeta}}\over{(\xi-\xi_+)(\xi-\xi_-)}}\eqno{(59)}$$ with $\xi_\pm=\pm(\mu-{{i\epsilon}\over{2\mu}})=\pm(\sqrt{3}-{{i\epsilon}\over{2\sqrt{3}}})$. According to $\zeta <0$ or $\zeta >0$, we integrate in the upper or lower complex $\xi$-plane, respectively obtaining $$\Delta_F(\zeta)^\prime=-{{i}\over{2\sqrt{3}}}e^{\pm i\sqrt{3}\zeta}.\eqno{(60)}$$ Continuing to $\zeta=0$, $$\Delta_F(0)^\prime=-{{i}\over{2\mu}}=-{{i}\over{2\sqrt{3}}},\eqno{(61)}$$ which, as will be shortly seen, provides a mass renormalization to the bare theory. 

\

To construct the generating functional $Z$ of the $n$-point correlation functions, one must to introduce a source factor $J(\zeta)$ to the Lagrangian $l$, defining $$l_J=-4(({{d\Omega}\over{d\zeta}})^2+\Omega^4-J\Omega).\eqno{(62)}$$ In order to take the theory to the standard ${{\lambda}\over{4!}}\Phi^4$ form, where $\lambda$ is the bare coupling constant, we redefine again the field function as $$\omega(\zeta)=2\sqrt{2}\Omega(\zeta),\eqno{(63)}$$ and the source function, $$j(\zeta)=\sqrt{2}J(\zeta),\eqno{(64)}$$ obtaining the action $$\sigma[\omega,j]=\int d\zeta(l_0+l_{int}+j\omega)\eqno{(65)}$$ with $$l_0=-{{1}\over{2}}(\partial_\zeta\omega)^2, \ l_{int}=-{{3/2}\over{4!}}\omega^4,\eqno{(66)}$$ i.e. $\lambda=3/2$. 

\

Then $Z[j]$ is given by$^7$ $$Z[j]={{exp(i\int d\zeta \ l_{int}(-i{{\delta}\over{\delta j(\zeta)}}))Z_0[j]}\over{exp(i\int d\zeta^\prime \ l_{int}(-i{{\delta}\over{\delta j(\zeta^\prime)}}))Z_0[j]}\vert_{j=0}},\eqno{(67)}$$ where $$Z_0[j]=exp(-{{i}\over{2}}\int d\zeta\int d\zeta^\prime j(\zeta)\Delta_F(\zeta-\zeta^\prime)^\prime j(\zeta^\prime))\eqno{(68)}$$ is the generating functional of the free theory i.e. that with $\lambda=0$. 

\

The $n$-point correlation functions are given by $$\pi(\zeta_1,\dots,\zeta_n)=<vac\vert\Pi(\hat{\omega}(\zeta_1)\dots\hat{\omega}(\zeta_n))\vert vac>=(-i)^n{{\delta^nZ[j]}\over{\delta j(\zeta_1)\dots\delta j(\zeta_n)}}\vert_{j=0},\eqno{(69)}$$ where $\Pi$ denotes the path ordered product of the quantum field operators $\hat{\omega}(\zeta_k)$ associated with the classical fields $\omega(\zeta_k)$, for $k=1,\dots,n$. The normalization $$Z[0]=1,\eqno{(70)}$$ i.e. the presence of the denominator in (67), guarantees the cancellation of vacuum diagrams in $\pi(\zeta_1,\dots,\zeta_n)$ in the formal perturbation expansion of the exponential function in (67). 

\

{\bf 6. Mass and coupling constant renormalization}

\

We shall not proceed to the derivation of the diagrams involved in the perturbation series, since the result is completely analogous to the usual spacetime ${{\lambda}\over{4!}}\Phi^4$ theory, where in (69) $\pi$ is replaced by $\tau$ and $\Pi$ by the time ordered product operator $T$. Instead, we want to emphasize three points, which suggest a possible further generalization to conformally flat metrics other than the four dimensional AdS case:

\

i) The quantum fluctuation of the classical metric hides a $\Phi^4$-theory in a one dimensional spacetime.

\

ii) The $\Phi^4$-theory is massless at the classical limit but, at the quantum level, it suffers a finite mass renormalization generated by the cosmological constant in the AdS spacetime, which in the $\Phi^4$-theory plays the role of a natural IR cutoff. In particular, to first order in $\lambda$, from (66), (67), (68) and (69), for the 2-point function one obtains $$\pi(\zeta_1,\zeta_2)=i\int_{-\infty}^{+\infty}{{d\xi}\over{2\pi}}{{e^{-i\xi(\zeta_1-\zeta_2)}}\over{\xi^2-m_r^2+i\epsilon}},\eqno{(71)}$$ where $$m_r^2=\delta m^2={{i\lambda}\over{2}}\Delta_F(0)^\prime={{\lambda}\over{4\mu}}={{\lambda}\over{4a\sqrt{-\Lambda}}}={{\sqrt{3}}\over{8}}.\eqno{(72)}$$

iii) The coupling constant of the scalar field is also finitely renormalized by the IR cutoff $\mu^2$. In fact, to second order of the perturbation expansion, the connected 4-point function in momentum space is given by $$\pi_4(p)=-i\lambda+3(-i\lambda)^2I(p)\eqno{(73)}$$ where $$I(p)={{1}\over{2\pi}}\int_{-\infty}^{+\infty}{{d\xi}\over{(\xi^2-\mu^2+i\epsilon)((p-\xi)^2-\mu^2+i\epsilon)}}=\int_0^1{{dt}\over{2\pi}}\int_{-\infty}^{+\infty}{{d\xi}\over{(\xi^2-\mu^2+p^2t(1-t)+i\epsilon)^2}},\eqno{(74)}$$ with $p=p_1+p_2$ ($p_i$, $i=1,2$: ingoing momenta).$^8$ Then $$I(0)={{1}\over{2\pi}}\int_{-\infty}^{+\infty}{{d\xi}\over{(\xi^2-\mu^2+i\epsilon)^2}}={{1}\over{2\pi}}\int_{-\infty}^{+\infty}{{d\xi}\over{(\xi-\xi_+)^2(\xi-\xi_-)^2}}:=\matrix{Lim\cr \zeta\to 0\cr} I(0,\zeta),\eqno{(75)}$$ with $$I(0,\zeta)= {{1}\over{2\pi}}\int_{-\infty}^{+\infty}{{d\xi e^{-i\xi\zeta}}\over{(\xi-\xi_+)^2(\xi-\xi_-)^2}},\eqno{(76)}$$ and $\xi_\pm=\pm(\mu-{{i\epsilon}\over{2\mu}})=\pm(\sqrt{3}-{{i\epsilon}\over{2\sqrt{3}}})$. Making the complex integrations as for the case of $\Delta_F(\zeta)^\prime$, one obtains $$I(0)={{i}\over{4\mu^3}}.\eqno{(77)}$$ So, the renormalized coupling constant results $$i\pi_4(0)=\lambda_r=\lambda(1+{{3\lambda}\over{4\mu^3}})=\lambda(1+{{3\lambda}\over{4a^3(-\Lambda)^{3/2}}})={{3}\over{2}}(1+{{\sqrt{3}}\over{8}}).\eqno{(79)}$$ 

\

{\bf 7. Final comments}

\

It is interesting to see the natural appearance of a massless scalar field derived from a conformally flat metric in four dimensions via a Weyl transformation. This unexpected result suggests several questions which could serve as motivation for further research. Firstly, since this result is valid for all conformally flat metrics in four spacetime dimensions, one could wonder what the consequences of the appearance of a massless scalar field are for all Robertson-Walker geometries, which have all been shown to be conformally flat$^9$ and which describe the large-scale structure of a homogeneous and isotropic Universe. Could this massless scalar field be interpreted as a natural-appearing inflation field? If so, what are the implications of this fact for the field of Quantum Cosmology? Secondly, is it possible to obtain a similar result in higher dimensional conformally flat spacetimes within the framework of General Relativity? In particular, can one construct a quantum theory of such a field in five dimensional Anti-De Sitter spacetime? This would be an interesting question related to the well known AdS/CFT correspondence, and which could help physicists to better understand this duality. And finally, can one obtain a similar higher dimensional result considering alternative theories of gravity such as Einstein-Gauss-Bonnet gravity or Lovelock theories? We leave all these questions open to further research.

\

{\bf References}

\

1. Romero, C., Fonseca-Neto, J.B., and Pucheu, M.L. (2012). Conformally Flat Spacetimes and Weyl Frames, Foundations of Physics {\bf 42}, 224-240. arXiv:1101.5333v1 [gr-qc].

\

2. Carroll, S. (2004). {\it Spacetime and Geometry. An Introduction to General Relativity}, Addison Wesley, San Francisco; p.130.

\

3. d'Inverno, R. (1992). {\it Introducing Einstein's Relativity}, Oxford University Press, Oxford; p.89.

\

4. Narlikar, J.V., and Padmanabhan, T. (1983). Quantum cosmology via path integrals, Physics Reports {\bf 100}, 151-200; p.164.

\

5. Schulman, L.S. (1981). {\it Techniques and Applications of Path Integration}, J. Wiley, New York; pp.88,95,96.

\

6. L. H. Ryder. (1996). {\it Quantum Field Theory}, Cambridge University Press, Cambridge; p.184.

\

7. Ref. 6, p.200.

\

8. Ref. 6, p.315.

\

9. M. Ibison. (2007). On the conformal forms of the Robertson-Walker metric, Journal of Mathematical Physics {\bf 48}, 122501/1-23. arXiv: 0704.2788 [gr-qc].

\

\

\

$^*$ {\it Postdoctoral fellowship, CONICET, Argentina}

\

$^{**}$ {\it With a leave of absence from Instituto de Ciencias, Universidad Nacional de General Sarmiento, Argentina}

\

e-mails: hugo.camargo@correo.nucleares.unam.mx, egurreta@ungs.edu.ar, socolovs@nucleares.unam.mx

\end

\centerline{O. Brauer}
\centerline{\it Facultad de Ciencias, Universidad Nacional Aut\'onoma de M\'exico}
\centerline{\it Circuito Exterior, Ciudad Universitaria, 04510, M\'exico D. F., M\'exico} 

\centerline{E. Kirchuk}
\centerline{\it Departamento de Matem\'aticas, CBC, Universidad de Buenos Aires}
\centerline{\it Ciudad Universitaria, Av. Cantilo S/N, Pab. III, CP 1428, Ciudad de Buenos Aires, Argentina}

\centerline{L. Raviola}
\centerline{\it Instituto de Industria, Universidad Nacional de General Sarmiento}
\centerline{\it J. M. Guti\'errez 1153, CP 1630, Los Polvorines, Pcia. de Buenos Aires, Argentina}

\centerline{M. Socolovsky}
\centerline{\it  Instituto de Ciencias Nucleares, Universidad Nacional Aut\'onoma de M\'exico}
\centerline{\it Circuito Exterior, Ciudad Universitaria, 04510, M\'exico D. F., M\'exico} 

\

{\bf Abstract.} {\it We compute the entropy of a Rindler particle-detector (observer) in the presence of a quantum field in the Minkowski vacuum state; due to the Unruh effect, the observer is immersed in a thermal bath at a temperature proportional to its proper acceleration. }

\

{\bf 1.} Hyperbolic motion [1] is the simplest non trivial motion of a classical massive particle in the context of special relativity. It consists, in the instantaneous rest frame, of a rectilinear accelerated motion with constant (proper) acceleration $\alpha$. Restricting ourselves for simplicity to the $(ct,x)$-plane of Minkowski spacetime, the trajectory of a particle which at $t=0$ passes through the point $(ct,x)=(0,c^2/\alpha)$ with zero velocity is given by $$x^2-c^2t^2=(c^2/\alpha)^2 \eqno{(1)}$$ with velocity $$\dot{x}={{\alpha t}\over{\sqrt{1+(\alpha t/c)^2}}}.\eqno{(2)}$$ It is clear that the lines $x=\pm ct$, respectively are the future and past horizons for the particle: no signal can reach to it if emitted at $(ct,x)$ with $ct>x$, and the particle can not send information to any point $(ct,x)$ with $ct<x$. (See Fig. 1.)

\

\centerline{\epsfxsize=65ex\epsfbox{1.eps}}

\

\centerline{Fig. 1}

\

The motion extends from $x=+\infty$ at $t=-\infty$ to $x=+\infty$ at $t=+\infty$. Asymptotically, the velocity tends to $-c$ and $+c$ respectively. The right region interior to the horizons is called the right Rindler wedge or, more simply, Rindler space; it is denoted by ${\cal R}$ [2]. As $\alpha \to +\infty$, the hyperboles tend to the horizons, so ``the proper acceleration of a photon can be taken to be infinite" [3]; while as $\alpha \to 0_+$, $c^2/\alpha\to +\infty$ and the particle is at rest at $x=+\infty$ (dashed line in Fig. 1). ${\cal F}$, ${\cal L}$, and ${\cal P}$ in Fig. 1 are the future, left, and past wedges. The relation between proper time $\tau$ and coordinate time $t$ is $$t={{c}\over{\alpha}}Sh({{\alpha\tau}\over{c}}),\eqno{(3)}$$ and the acceleration vector is $$\alpha^\mu=(\alpha^0,\alpha^1)={u^\mu}^\prime=\gamma\dot{u}^\mu=\gamma(c\dot{\gamma},{{d}\over{dt}}(\gamma\dot{x}))\eqno{(4)}$$ with $u^\mu=(c\gamma,\gamma\dot{x})$, $\gamma=1/\sqrt{1-{{\dot{x}^2}\over{c^2}}}$, ${u^\mu}^\prime={{du^\mu}\over{d\tau}}$, $\dot{u}^\mu={{du^\mu}\over{dt}}$. From $$-\eta_{\mu\nu}\alpha^\mu\alpha^\nu=\alpha^2\eqno{(5)}$$ with $\eta_{\mu\nu}=\pmatrix{1 & 0 \cr 0 & -1\cr}$, one obtains $$\alpha^1=\sqrt{\alpha^2+{{1}\over{c^2}}{{(\dot{x}\ddot{x})^2}\over{(1-\dot{x}^2/c^2)^4}}}=\alpha Ch({{\alpha\tau}\over{c}}).\eqno{(6)}$$

\

{\bf 2.} In 1976, Unruh [4] found that, if a quantum field in its vacuum state is present in the Minkowski spacetime, the accelerated particle (``Rindler observer" or ``particle-detector") would detect that vacuum as a thermal bath or radiation with absolute temperature $$T={{\hbar\alpha}\over{2\pi k_Bc}} \ (={{\alpha}\over{2\pi}} \ in \ natural \ units)\eqno{(7)}$$ with $\hbar$ and $k_B$ the Planck and Boltzmann constants. This means that to have a temperature of $1 \ ^\circ K$ one needs an acceleration of approximately $10^{18}{{m}\over{sec^2}}$! For the light cones $T\to +\infty$, and for a particle at rest or in uniform motion $T=0$. Assuming with Unruh that the accelerated body comes to equilibrium with the radiation [5], one can consider each hyperbolic motion as a {\it reversible isotherm} at temperature $T$. 

\

{\bf 3.} In 1987, Laflamme computed the entropy of the Rindler wedge ${\cal R}$, and found an infinite result: one quarter of the area of the event horizon (similarly to what occurs for the Schwarzschild black hole [7]). The claim was, however, that the entropy per unit area is finite. The result fits together with the infinite value of the temperature at the horizon; however, it says nothing with respect to an entropy $S(T)$ which can be naturally associated to each accelerated motion, corresponding to a temperature $T$. This leads to the possibility of studying some aspects of the thermodynamics of the particle-detector.

\

{\bf 4.} Our starting point is the relation between the variation of internal energy, absorbed heat, and work done on a system, as given by the first and second laws of thermodynamics: $$dU=\delta Q+\delta W.\eqno{(8)}$$ For a reversible process, $\delta Q=TdS$, and $\delta W=f^1dx$, where $f^1=m\alpha^1$ ($m$ is the mass of the particle). The internal energy is, in this case, the kinetic energy $E=\sqrt{p^2c^2+m^2c^4}$, with $p={{m\dot{x}}\over{\sqrt{1-\dot{x}^2/c^2}}}$, which gives $$E=mc^2\sqrt{1+({{\alpha t}\over{c}})^2}=mc^2Ch({{\alpha\tau}\over{c}}).\eqno{(9)}$$ Then, $$T{{dS}\over{dt}}={{dE}\over{dt}}-f^1\dot{x}, \eqno{(10)}$$ and using ${{d}\over{dt}}={{1}\over{Ch({{\alpha\tau}\over{c}})}}{{d}\over{d\tau}}$, one obtains $${{dS}\over{d\tau}}={{mc\alpha}\over{T}}Sh({{\alpha\tau}\over{c}})(1-Ch({{\alpha\tau}\over{c}}))={{2\pi k_Bmc^2}\over{\hbar}}Sh({{2\pi k_B}\over{\hbar}}T\tau)(1-Ch({{2\pi k_B}\over{\hbar}}T\tau))$$ $$(=2\pi mSh(2\pi T\tau)(1-Ch(2\pi T\tau)) \ in \ natural \ units).\eqno{(11)}$$ Integrating this expression between $\tau_1$ and $\tau_2$, the variation of the entropy along the isotherm is $$\Delta S(\tau_2,\tau_1;T)=S(\tau_2,T)-S(\tau_1,T), \eqno{(12)}$$ with $$S(\tau_i,T)={{mc^2}\over{T}}(Ch({{2\pi k_B}\over{\hbar}}T\tau_i)-{{1}\over{2}}Sh^2({{2\pi k_B}\over{\hbar}}T\tau_i))$$ $$(={{m}\over{T}}(Ch(2\pi T\tau_i)-{{1}\over{2}}Sh^2(2\pi T\tau_i)) \ in \ natural \ units), \ i=1,2.\eqno{(13)}$$ Since $$S(-\tau_i,T)=S(\tau_i,T),\eqno{(14)}$$ then $$\Delta S(\tau,-\tau;T)=0;\eqno{(15)}$$ in particular $\Delta S(-\infty,+\infty;T)=0$: for any finite part of the motion from $-\tau$ to $+\tau$ (or for the infinite total motion from $\tau=-\infty$ to $\tau=+\infty$) the isothermal completes a {\it cycle}, and the entropy, being a state function, returns to its initial value. 

\

In particular, $$\Delta S(\tau,0;T)={{mc^2}\over{T}}(Ch({{2\pi k_B}\over{\hbar}}T\tau)-{{1}\over{2}}Sh^2({{2\pi k_B}\over{\hbar}}T\tau)-1).\eqno{(16)}$$

\

For any fixed $T>0$, one can verify the following limit behavior: 

\

$\tau_2$ fixed, $\tau_1\to -\infty$ (past horizon): $$\Delta S(\tau_2,\tau_1;T)\sim {{mc^2}\over{8T}}e^{2\vert\lambda\vert}\to +\infty, \ \lambda={{2\pi k_B}\over{\hbar}}T\tau_1, \eqno{(17)}$$ maximal entropy, minimum order;

\

$\tau_1$ fixed, $\tau_2\to +\infty$ (future horizon): $$\Delta S(\tau_2,\tau_1;T)\sim -{{mc^2}\over{8T}}e^{2\eta}\to -\infty, \ \eta={{2\pi k_B}\over{\hbar}}T\tau_2, \eqno{(18)}$$ minimal entropy, maximum order. 

\

If we call $$h(y)=Ch(y)-{{1}\over{2}}Sh^2(y), \ y={{2\pi k_B}\over{\hbar}}T\tau, \eqno{(19)}$$ then $$\Delta S(\eta,\lambda;T)={{mc^2}\over{T}}(h(\eta)-h(\lambda))\eqno{(20)}$$ which, for $\lambda=\lambda_0=0$ ($\tau_1=0$) and $\eta>0$ ($\tau_2>0$) has the behavior plotted in Fig. 2.

\

\centerline{\epsfxsize=70ex\epsfbox{2.eps}}

\

\centerline{Fig. 2}

\

For arbitrary $\tau>0$, the expansion of $Ch(\eta)-{{1}\over{2}}Sh^2(\eta)-1$ near $\eta=0$ ($T=0$) leads to\ $$\Delta S(\tau,0;T)\cong -2\pi^4{{c^2k_B^4}\over{\hbar^4}}m\tau^4T^3+O(T^4).\eqno{(21)}$$ So, $\Delta S\to 0_-$ as $T\to 0_+$, i.e. the entropy remains constant for the inertial particle. In the other extreme, for large $T$ but fixed $\tau >0$ i.e. near the future horizon, $$\Delta S(\tau,0;T)\cong -{{mc^2}\over{8T}}e^{4\pi{{k_B}\over{\hbar}}T\tau}\to -\infty \ \ as \ \ T\to +\infty.\eqno{(22)}$$

\

In terms of the coordinate time $t$, from (3), (13) is given by $$S(t,T)={{mc^2}\over{T}}(\sqrt{1+{{4\pi^2k_B^2}\over{\hbar^2}}t^2T^2}-{{2\pi^2k_B^2}\over{\hbar^2}}t^2T^2),\eqno{(23)}$$ and so $$\Delta S(t,0;T)={{mc^2}\over{T}}(\sqrt{1+{{4\pi^2k_B^2}\over{\hbar^2}}t^2T^2}-{{2\pi^2k_B^2}\over{\hbar^2}}t^2T^2-1).\eqno{(24)}$$

\

{\bf 5.} One can give a geometric interpretation of the result (16) analogous to that given by Laflamme to the Rindler wedge, but different in the sense that now it is not the horizon area the relevant quantity. In the present case, the only spatial geometric characteristic of the hyperbolic motion is the spatial length $L$  traveled  by the observer during some interval of its proper time, say between 0 and $\tau$. An easy calculation leads to $$L={{c^2}\over{\alpha}}(Ch({{\alpha\tau}\over{c}})-1)\eqno{(25)}$$ and therefore, in terms of $L$ and $T$, (16) becomes $$\Delta S(L,0;T)=-2\pi^2({{k_B}\over{\hbar}})^2mL^2T\eqno{(26)}$$ which, per unit mass, in terms of the acceleration, and in natural units gives $${{\Delta S(L,0;T)}\over{m}}=-\alpha A,\eqno{(27)}$$ where $A=\pi L^2$. Though $A$ is the area of a disk, we can not at the moment give to it a physical interpretation. Another sensible area element in the present context is the shaded area in Fig. 1, which is given by $$A^\prime=({{c^2}\over{\alpha}})^2(1+{{\alpha L}\over{c^2}})ArcCh(1+{{\alpha L}\over{c^2}})-{{Lc^2}\over{\alpha}}\sqrt{1+{{2c^2}\over{L}}},\eqno{(28)}$$ which however is different from $A$. As a comparison with the Schwarzschild black hole the radious of the disk (equatorial cross section) is the Schwarzschild radious.

\

{\bf Acknowledgments.} M.S. thanks for hospitality to the Instituto de Ciencias de la Universidad Nacional de General Sarmiento, Pcia. de Buenos Aires, Argentina, where part of this work was done. This work was partially supported by the project PAPIIT IN101711-2, DGAPA, UNAM, M\'exico.

\

{\bf References.}

\

[1] Landau, L. D. and Lifshitz, E. M. (1975). {\it The Classical Theory of Fields, Course of Theoretical Physics, Vol. 2}, Elsevier, Amsterdam; p. 24.

\

[2] Rindler, W. (1966). Kruskal Space and the Uniformly Accelerated Frame, {\it American Journal of Physics} {\bf 34}, 1174-1178.

\

[3] Rindler, W. (2006). {\it Relativity: Special, General, and Cosmological}, Oxford University Press; p. 71.

\

[4] Unruh, W. G. (1976). Notes on black-hole evaporation, {\it Physical Review D} {\bf 14}, 870-892.

\

[5] Unruh, W. G. (1992). Thermal bath and decoherence of Rindler spacetimes, {\it Physical Review D} {\bf 46}, 3271-3277.

\

[6] Laflamme, R. (1987). Entropy of a Rindler wedge, {\it Physics Letters B} {\bf 196}, 449-450.

\

[7] Hawking, S. W. (1975). Particle Creation by Black Holes, {\it Communications in Mathematical Physics} {\bf 43}, 199-220; erratum: ibid. {\bf 46}, 206.

\

\

\

\

\

\

\

\

\

\

\

\

\

\

\

\

\

\

\

\

e-mails: brauer@ciencias.unam.mx; kirchuk@df.uba.ar; lraviola@ungs.edu.ar; 

socolovs@nucleares.unam.mx

\end